\documentclass[twocolumn,pre,superscriptaddress]{revtex4}
\usepackage{graphicx}
\usepackage{graphics}
\usepackage{epsfig}
\usepackage{times}

\begin{document}

\title{Optimized Folding Simulations of Protein A}

\author{Simon Trebst}
\affiliation{Microsoft Research, Station Q, University of California, Santa Barbara, CA 93106}

\author{Ulrich H.E. Hansmann}
\affiliation{Department of Physics, Michigan Technological University, Houghton, MI 49931}
\affiliation{John-von-Neumann Institute for Computing, 
             Forschungszentrum J\"ulich, D-52425 J\"ulich, Germany}
\date{\today}

\begin{abstract}
We describe optimized parallel tempering simulations of the 46-residue 
B-fragment of protein A. Native-like configurations with a root-mean-square
deviation of $\approx 3$ \AA\  to the experimentally determined structure
(Protein Data Bank identifier 1BDD) are found. However, at biologically relevant
temperatures such conformations appear with only  $\approx 10$\% frequency
in our simulations.
Possible short comings in our energy function are discussed.
\end{abstract}


\maketitle

\section{Introduction}
 
Rational drug design or the pathology of  amyloid diseases are only two
problems whose  solutions require a detailed understanding
of the relation between chemical composition and structure
(and function) of proteins. Exploring this relationship through numerical 
simulations is a computationally hard problem. 
Two major factors limit our ability to efficiently simulate large proteins 
and study their folding transitions.
First, statistically sampling the rough energy landscape of a protein can be
extremely slow even at room temperature. Second, present energy functions 
are often insufficiently accurate in describing the interactions between the 
atoms within a protein, and between protein and their surrounding solvent.
It is often not clear whether the failure of a computer experiment to find the 
known structure of a protein results from poor sampling or lack of accuracy 
in the energy function. 

To overcome some of the limitations of statistical sampling in the simulation
of small proteins, sophisticated simulation schemes such as parallel tempering \cite{PT1,PT2}
or generalized ensemble methods \cite{HO96g,Oliveira} are now widely employed 
numerical methods \cite{KK07, Hamacher, Wei}
In a recent line of research feedback-optimized algorithms have been developed
that aim at further improving the statistical sampling of these methods by 
systematically improving the simulated statistical ensemble 
\cite{OptimizedEnsembles}, 
e.g. the exact placement of replicas in temperature space 
\cite{OptimizedTempering,TTH06,GK07}.
Recent simulations of the 36-residue villin headpiece sub-domain HP-36
in Ref.~\onlinecite{TTH06}  demonstrated that optimizing the sampled
temperature distribution leads to qualitatively different results for the
same force field, in this case a combination of the ECEPP/3 force field \cite{EC} 
with an implicit solvent \cite{OONS}. 
\begin{figure}[b]
  \includegraphics[width=0.75\columnwidth]{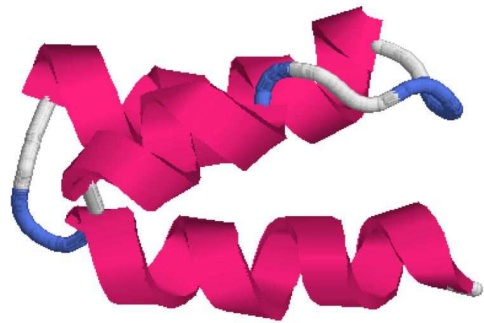}
  \caption{The experimentally determined structure of the 46-residue B-fragment of protein A
                  as stored in the Protein Data Bank (identifier 1BDD).
                  \label{1bdd}}
\end{figure}
Previous simulations of HP-36 in Ref.~\cite{LHH04} had indicated that the 
native structure is {\it not} the global free energy minimum at room temperature for
this force field. However, with the optimized temperature distribution it was later
found that the correct structure is sampled with $\approx 90$\% frequency at
room temperature \cite{TTH06}. Clearly, the earlier numerical simulations
suffered from a sampling problem and resulted in misleading conclusions on 
the force field, while the energy function in fact accurately described the 
interactions for HP-36 as shown in the latter study with improved sampling. 
While this result is promising, the observed sampling difficulties nevertheless 
suggest that the energy landscape for HP-36 modeled by this force field may
be more rough than the one experienced by the ``real'' protein.
Given the complexity of the energy landscape for this small protein, one would
expect that with increasing size and complexity of the molecule the accuracy of
the energy function will further decrease as the energy landscape gains further
complexity. This would render it even more difficult (or simply impossible) to pick the
correct structure in the numerical simulation of larger proteins. 
As stable domains in proteins usually consists of 50-200 residues it is therefore 
important to test the accuracy of current energy functions for proteins of this size. 
As a first step in this direction we have applied the feedback-optimized parallel
tempering scheme to simulate the  46-residue fragment 10-55 of the B domain of 
protein A (Protein Data Bank identifier 1BDD) which forms the three-helix 
bundle displayed in Fig.~\ref{1bdd} as determined in experiments \cite{BDD}.
This is one of the few small proteins with experimentally well-characterized states.
For this reason, it has raised some interest as a model to test folding algorithms
or energy functions \cite{Brooks,Ghosh,Irbaeck,Garcia,Wenzel,Unres,Shakhnovich}.
While our high-statistics simulations do find native-like configurations with a 
root-mean-square deviation of $\approx$ 3 \AA\  to the experimentally determined 
structure, these configurations do not correspond to the lowest energy configuration
and appear with only around 10\% probability at biologically relevant temperatures. 
Our results indicate that this deviation from experiment is due to a bias in the ECEPP/3
energy function toward helical structures. 
 Another contributing factor is the
use of  an implicit solvent model in our simulations. These models  reduce
dramatically the numerical costs of protein simulations  but can lead
to distorted free energy landscapes. This effect  has been observed earlier \cite{ZB02,NG03} 
 even for more  sophisticated implicit solvents than used by us. 

\section{Methods}
Our simulations of the protein A fragment utilize the ECEPP/3 force field \cite{EC} as
implemented in the 2005 version of the program package SMMP \cite{SMMP,SMMP05}.
Here, the interactions between the atoms within the protein are approximated
by a sum $E_{\text{ECEPP/3}}$ consisting of electrostatic energy $E_C$, 
a Lennard-Jones term $E_{LJ}$, a hydrogen-bonding term $E_{hb}$ 
and a torsion energy $E_{tor}$:
\begin{eqnarray}
  E_{\text{ECEPP/3}} &=& E_C + E_{LJ}  + E_{hb} + E_{tor} \nonumber \\
  &=&  \sum_{(i,j)} \frac{332 q_i q_j}{\epsilon r_{ij}} \nonumber \\
 &&   + \sum_{(i,j)} \left( \frac{A_{ij}}{r_{ij}^{12}} - \frac{B_{ij}}{r_{ij}^6} \right) \nonumber \\ 
 &&   + \sum_{(i,j)} \left( \frac{C_{ij}}{r_{ij}^{12}} - \frac{D_{ij}}{r_{ij}^{10}} \right) \nonumber \\ 
  && + \sum_l U_l ( 1\pm \cos(n_l \xi_l)) \;,
\end{eqnarray}
where $r_{ij}$ is the distance between the atoms $i$ and $j$,  $\xi_l$ is the $l$-th torsion 
angle, and energies are measured in kcal/mol.  
The protein-solvent interactions are approximated by a solvent accessible surface term 
\begin{equation}
  E_{solv} = \sum_i \sigma_i A_i \;.
\end{equation}
The sum goes over the  solvent accessible areas $A_i$ of all atoms $i$ weighted by solvation
parameters $\sigma_i$ as determined in Ref.~\onlinecite{OONS}, a common choice when the
ECEPP/3 force field is utilized.

\begin{figure}
    \includegraphics[width=\columnwidth]{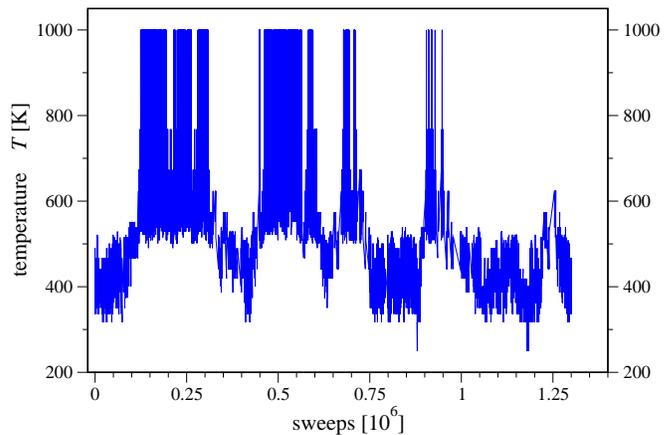}
    \caption{ Time series of visited temperatures for one of the replicas in the
                         parallel temperature simulation. Time is measured in MC sweeps and starts
                         with the begin of measurements.
                          \label{traject}}
\end{figure}

The above defined energy function leads to an energy landscape that is characterized by a 
multitude of minima separated by high barriers. In order to improve statistical sampling and
speed up equilibration at low temperatures we utilize a parallel tempering scheme\cite{PT1,PT2}
which was first used in protein science in Ref.~\cite{H97f}. 
In this scheme $N$ non-interacting copies, or ``replicas'', of the protein are simultaneously 
simulated at a range of temperatures $\{T_1, T_2, \ldots, T_N\}$. After a fixed number of 
Monte Carlo sweeps (or a molecular dynamics run for a fixed time interval) a sequence of 
swap moves, the exchange of two replicas at neighboring temperatures, $T_i$ and $T_{i+1}$, 
is suggested and accepted with a probability
\begin{equation}
  p(E_i, T_i \rightarrow E_{i+1}, T_{i+1}) = \min \left( 1,  \exp(\Delta\beta \Delta E) \right)\;,
  \label{Eq:SwapAcceptance}
\end{equation}
where $\Delta\beta = 1/T_{i+1} - 1/T_i$ is the difference between the inverse temperatures 
and $\Delta E = E_{i+1} -E_i$ is the difference in energy of the two replicas. 
The exchange of conformations considerably improves equilibration for all replicas, 
especially those at low temperatures which can have extremely long equilibration times
for conventional canonical simulations (at a fixed temperature).

\begin{figure}
    \includegraphics[width=\columnwidth]{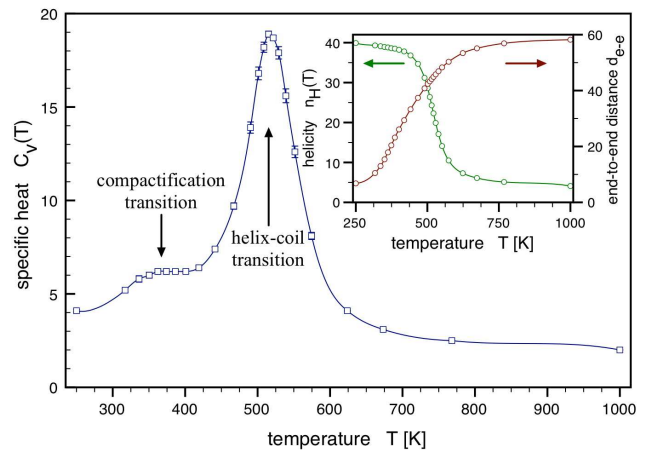}
    \caption{The specific heat $C(T)$ as function of temperature $T$. A peaked signal
                          is found at the helix-coil transition around $T_c = 515~K$ where the 
                          helicity $n_H(T)$ increases, see the inset. 
                          Below the transition a shoulder emerges over a broad temperature regime 
                          in which the average end-to-end distance $d_{e-e}$ of the protein 
                          decreases rapidly as shown in the inset, 
                          an indication that the helices form a secondary structure.
                          \label{spec_heat}}
\end{figure}

The improved equilibration of the replica exchange scheme is due to the random walks 
that individual replicas can perform in in temperature space allowing them to move to
higher temperatures where equilibration is fast and then move back down to lower
temperatures thereby escaping  barriers in the energy landscape.
Obviously, the number of round-trips $n_{rt}$ between the lowest and highest temperature, 
$T_1$ and $T_N$, respectively, is a lower bound for the  statistically independent visits at 
the lowest temperature, and therefore a good measure for the equilibration of the parallel 
tempering simulation.
It is therefore desirable to maximize the number of round trips by optimizing the position
of temperature points in the interval $[T_1, T_N]$. This can indeed be achieved in a 
systematic way by feeding back the local diffusivity of the described random walk using
the feedback algorithm described in Refs.~\onlinecite{OptimizedTempering,TTH06}. 
Technically, replicas are ``labeled" according to which of the two extremal temperatures,
$T_1$ or $T_N$, they have visited last. Using this label we can define the number of
replicas $n_{\rm up}(i)$ ($n_{\rm down} (i)$) which at temperature $T_i$ have come
from $T_1$ ($T_N$).
Then the fraction of replicas moving in one direction
\begin{equation}
    f_{\rm up} (i) = \frac{n_{\rm up}(i)}{n_{\rm up}(i) + n_{\rm down}(i)}
\end{equation}
describes a stationary distribution of probability flow between temperatures $T_1$ and 
$T_N$  with boundary conditions $f_{\rm up} (1)  = 1$ and $f_{\rm up} (N) = 0$. The local
diffusivity $D(T)$ of the random walk which a single replica performs in temperature is
then given by $D(T) \propto \Delta T \cdot \left( df/dT \right)^{-1}$, where $\Delta T$ is the
size of the temperature interval around $T$.
Feeding back this local diffusivity it was shown in Refs.~\onlinecite{OptimizedTempering,TTH06}
that for the optimal temperature distribution, i.e. the one that maximizes the number of round trips
$n_{RT}$, the fraction decreases linearly
\begin{equation}
f^{\rm (opt)}_{\rm up} (i) = \frac{N-i}{N-1}~.
\end{equation}
For our simulations of the protein A fragment such an optimized distribution can be found 
by applying the iterative procedure described in Refs.~\onlinecite{OptimizedTempering,TTH06} and 
is given for the 24 replicas in our simulations by
$1000$, $768$, $673$, $624$, $574$, $552$, $539$, $529$, $521$, $515$, $509$, $501$, 
$490$, $467$, $441$, $419$, $401$, $386$, $374$, $362$, $350$, $336$, $317$, $250$. 
Our data are taken from a simulation with 1,000,000 sweeps for each replica. 
Here, a sweep consists of a sequential series of attempts to update each of the 276 dihedral 
angles (the true degrees of freedom in our model) once. After each sweep, we attempt an
exchange move (swap) of configurations between neighboring temperatures which is
accepted with probability (\ref{Eq:SwapAcceptance}). 
The walk of one replica along the ladder of temperatures is displayed in Fig.~\ref{traject}.
Measurements are taken every ten sweeps and stored for further analysis.
These include the energy $E$,  the radius of gyration $r_{gy}$ as a measure of the geometrical size, 
and the number of helical residues $n_H$, i.e. residues where the pair of dihedral angles $(\phi,\psi)$ 
takes  values in the range ($ -70^\circ \pm 30^\circ $, $ -37^\circ \pm 30^\circ $). 
Finally we recorded the configurations with overall lowest energy obtained in our simulation.

\section{Results and Discussions}
The biologically active state of a protein is thought to be the global minimum of the free energy at room
temperature. Heating leads to unfolding that is reversible after cooling. Hence, the folding
transition should be marked by a signal in the specific heat
\begin{equation}
    C(T) = \beta^2 (\langle E^2 \rangle - \langle E \rangle^2)/ N
  \end{equation} 
 with $\beta = 1/k_B T$ ($k_B$ is the Boltzmann constant) and $N$ the number of residues.
For protein A we indeed  find a pronounced peak in the specific heat, 
displayed in Fig.~\ref{spec_heat}, at a temperature $T_c = 515$~K.  
This peak is related to the formation of $\alpha$-helices  as one can see from
the inset where the average number $n_H$ of residues which are part of an $\alpha$-helix is
plotted versus temperature.  Around the transition temperature $T_c$ the helicity rapidly increases. 
The corresponding formation of hydrogen bond between residues $(i,i+4)$ leads to a much 
lower energy of such configurations, and the resulting fluctuation in the average energy as 
function of temperature is measured by the specific heat. 
Below this helix-coil transition a shoulder is observed in the specific heat in a temperature
range between $330 - 430$~K. Since the helicity varies only little in this temperature range,
but the typical end-to-end distance $d_{e-e}$ of the protein A configurations decreases rapidly
-- as shown in the  inset of Fig.~\ref{spec_heat} -- this temperature regime is marked by the
formation of a secondary structure of the helical segments which we will discuss in detail below.
As the temperature is further decreased below $T \approx 330$~K, the specific heat lowers 
again and the end-to-end distance approaches a constant value.

\begin{figure}
    \includegraphics[width=\columnwidth]{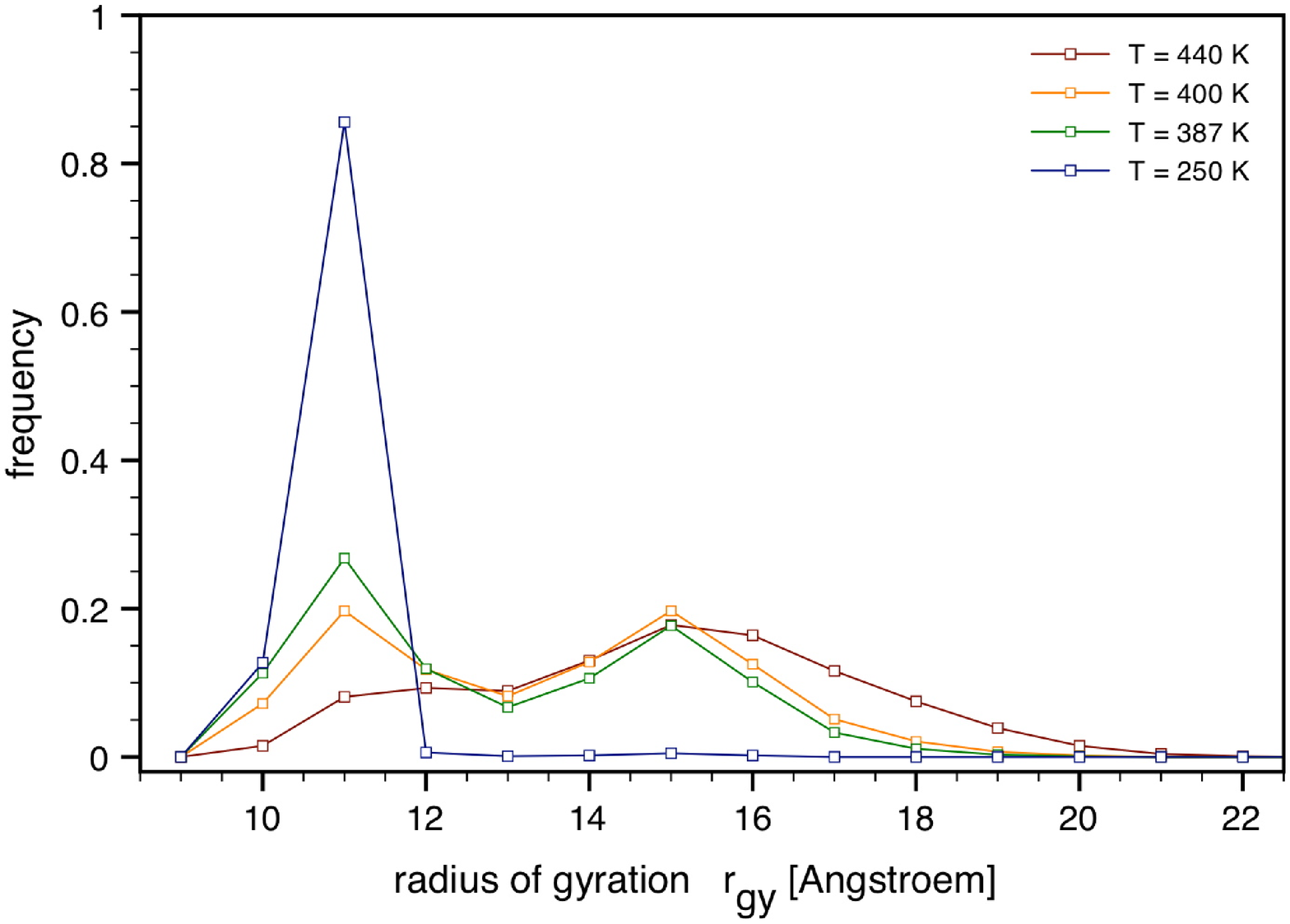} 
    \includegraphics[width=\columnwidth]{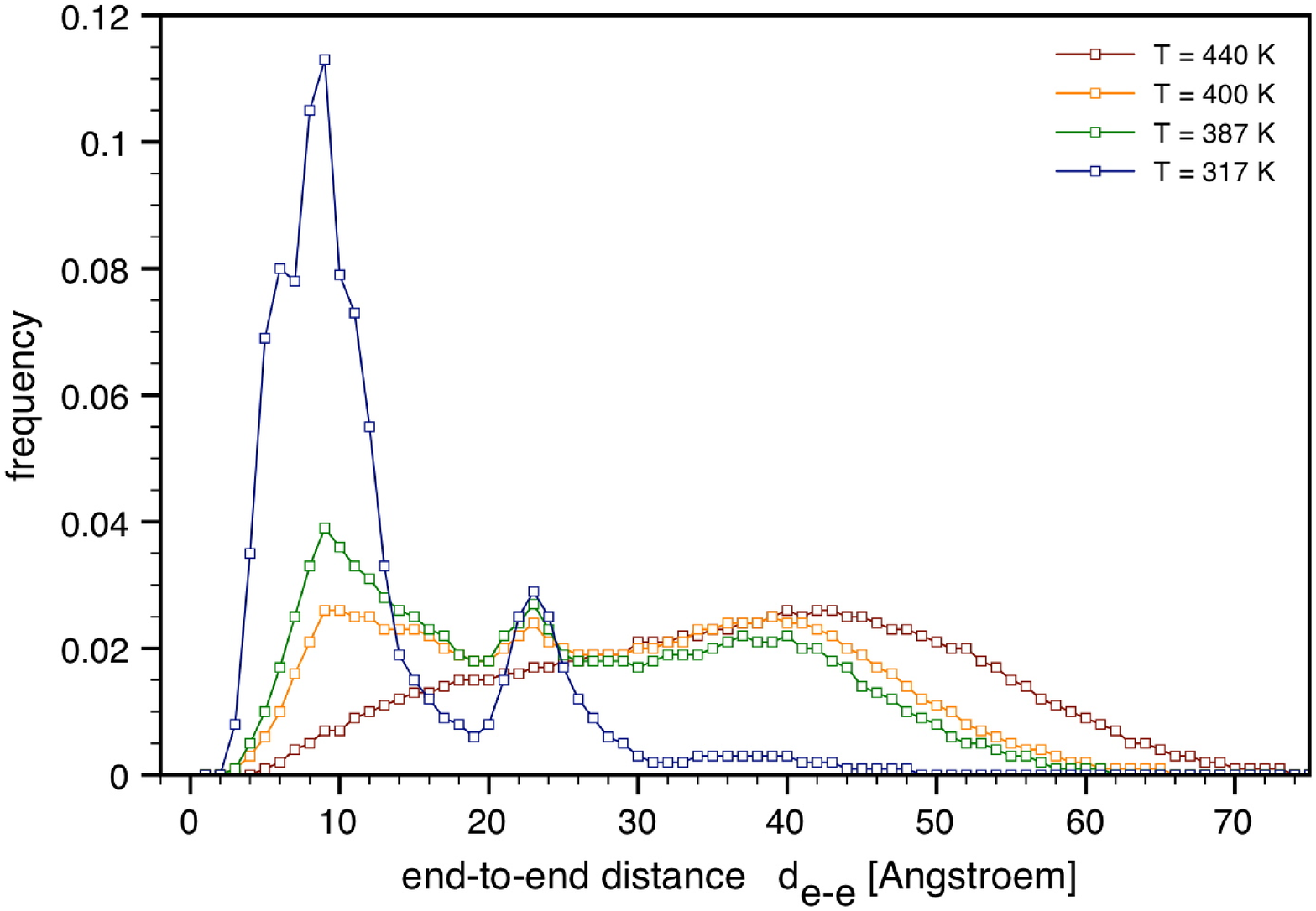} 
    \caption{Frequency of (a) the radius of gyration $r_{gy}$  and (b) the
                         end-to-end distance $d_{e-e}$ for various temperatures.
                         Each data point samples entries of a bin with 1 \AA\  width. 
                         The histograms are normalized.
                         \label{histograms}} 
\end{figure}

\begin{figure}[t]
    \includegraphics[width=0.525\columnwidth]{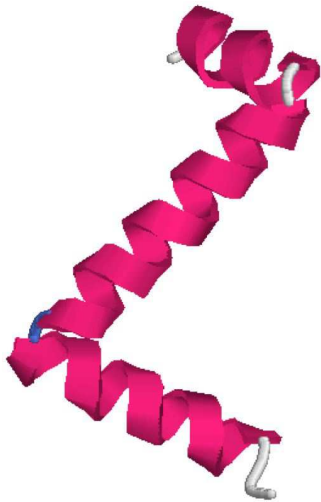} 
    \caption{Typical extended low-energy structure with high helical content (Type I).
                     \label{structures_a} }
\end{figure}

A more detailed picture of the formation of a compact secondary structure emerges when
looking at the frequency of configurations with typical end-to-end distances $d_{e-e}$ and their
respective radius of gyration $r_{gy}$, a measure for the compactness of a folded protein
structure, which are both plotted versus temperature in Fig.~\ref{histograms}.
At temperatures directly below the helix coil transition, e.g. our data point at $T = 440$~K,
the histograms indicate that configurations still differ widely as seen from the broad 
distributions found in both histograms, but with a clear preference for extended structures
with a radius of gyration centered around $r_{gy} = 15$ \AA.
  In the shoulder region of the specific heat, e.g. at our data points $T= 400$~K and 
$T = 387$~K, a double peak appears in histogram of the radius of gyration indicating 
the competion between extended structures ($r_{gy} \approx 15$ \AA) and compact 
ones ($r_{gy} \approx 11$ \AA). 
Further decreasing the temperature we are finally left with compact configurations only
(with $r_{gy} \le 11$ \AA). The histograms in the radius of gyration seem to indicate
a compactification transition where distinct secondary structures are formed 
at around $387-400$~K that separates compact helical structures from extended 
helical structures immediately below the helix-coil transition around $T_c = 515$~K. 
Indeed, we find that two different compact structures are primarily formed below this
compactification transition as becomes evident from our measurements of frequencies of
typical end-to-end distance for various temperatures. 
Above the compactification transition and below the helix-coil transition the 
histogram of typical  end-to-end distances is first centered around $d_{e-e} = 43$ \AA\ 
for $T=440$~K and an almost flat histogram is observed at $T= 400$~K for 
$d_{e-e} = 9-40$ \AA.
Below $T= 387$~K two additional peaks form around $d_{e-e} = 9$ \AA\  and 
$d_{e-e} = 23$ \AA, while there is still a broad feature around $d_{e-e} = 40$ \AA.
Further lowering the temperature, the two peaks $d_{e-e} = 9$ \AA\  and $d_{e-e} = 23$ \AA\ 
further proliferate and become the dominant feature in the histogram.
Finally, towards the lowest temperature $T=250$~K which we sampled, only configurations 
with small end-to-end distance prevail with more than 90\% of all sampled configurations 
having a typical end-to-end distance of $d_{e-e} \le 10$ \AA.

\begin{figure}[t]
    \includegraphics[width=0.75\columnwidth]{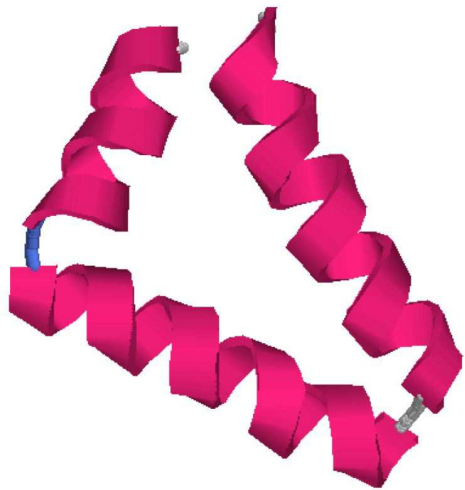} 
    \caption{One of the typical triangle shaped compact structures characterized by small 
                    end-to-end distance $d_{e-e}$ (Type II).
                    \label{structures_b} }
\end{figure}

\begin{figure}[t]
    \includegraphics[width=0.75\columnwidth]{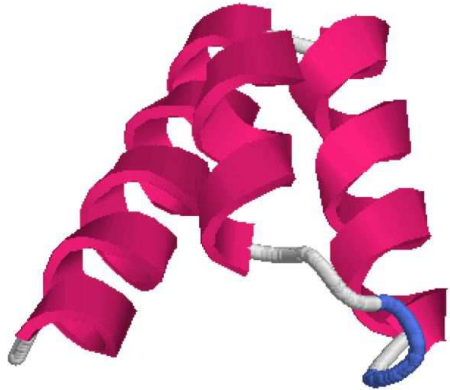} 
    \caption{Another typical compact structure, but with larger end-to-end 
			   distance $d_{e-e}$ (Type III). 
			   This structure resembles the native structure as deposited in 
			   the PDB and shown in Fig.~1.
			   \label{structures_c} } 
\end{figure}

 The above analysis of the histograms of typical radius of gyration and end-to-end distance
indicate that below the helix-coil transition temperature of $T_c =515$~K there exist 
{\em three} different types of configurations which all have high helix-content ($n_H \approx  32-40$)
but differ strongly in their respective arrangement of the helical segments. 
Two of these structures are compact, while one is extended and found only above the 
compactifcation transition around $T \approx 387$~K.
    A typical example for the extended helical structure -- named  ``structure I'' -- is shown in 
Fig.~\ref{structures_a}. The illustrated configuration has a radius of gyration of $r_{gy} = 13.5.$ \AA,
an end-to-end distance of $d_{e-e} = 34.0$ \AA, and a solvent-accessible surface area of
$A_{\rm SAS} = 4440$ \AA$^2$.  All three helical segments are formed, but the total helicity
is with 39 helical residues higher than the one found for the PDB structure where
only 34 residues are part of an $\alpha$-helix. Accordingly, its root-mean square deviation 
(rmsd) to the native structure (as deposited in the Protein Data Bank under the identifier 1BDD) 
is 11.5 \AA\ (calculated over backbone atoms only).  
The total energy after minimization is $E_{\rm tot} = -643$~kcal/mol, of which 
$E_{solv} = -186$~kcal/mol result from the solvation energy and 
$E_{\text{ECEPP/3}}=-457$~kcal/mol from the intramolecular interactions. 
   Fig.~\ref{structures_b} displays the first of the two compact structures found below the 
compactification transition. Named ``structure II''  it is the configuration with lowest overall energy 
found in our extensive simulations. After minimization we have $E_{\rm tot} = -692$~kcal/mol, 
with the energy difference to the previous structure resulting from the  more favorable intramolecular 
interactions ($E_{\text{ECEPP/3}} = -510$~kcal/mol)  while there is little difference in the solvation term, 
$E_{solv} = -182$~kcal/mol. The small end-to-end distance of $d_{e-e} = 5.7$ \AA\  reflects a 
{\rm triangle shape} of this configuration which differs from the native one by a rmsd of 7.5 \AA\ 
and  is  with 41 residues maximal helical. The configuration is with $r_{gy} = 11.3$ \AA\ and a
a solvent accessible surface area $A_{SAS} = 3950$ \AA$^{2}$ more compact than the
extended structure (type I) but not as compact as the native one which has  $r_{gy} = 9.7$ \AA,   
and a solvent-accessible surface area $A_{\rm SAS} = 3333$ \AA$^2$. 
  A closer resemblance to the native structure has the second type of compact configurations found 
in our simulations, named ``structure III" and illustrated in Fig.~\ref{structures_c}. 
This structure differs from the other compact, triangle shaped structure (`` structure II'')  in that it 
is more compact but has a smaller helicity with only 37 residues which are  part of an $\alpha$-helix.  
For this configuration we have measured a radius of gyration of  $r_{gy}= 10.4$ \AA, 
a solvent-accessible surface area $A_{\rm SAS} = 3780$ \AA$^2$, and an end-to-end distance 
$d_{e-e} = 23.0$ \AA. Strikingly, the rmsd to the native structure is only 3.3 \AA.
However, the total energy $E_{tot} = -662$~kcal/mol is by about 30 kcal/mol 
higher than the one of the other compact structure shown in Fig.~\ref{structures_b}.  
This is similar to Ref.~\onlinecite{Shakhnovich} where the best sampled structure has a 
rmsd of 2.33 \AA\  to the native one, but the lowest energy configuration differs by
6.41 \AA. In our case, the
difference for the two compact structures is primarily due to the  ECEPP/3 energy
which for the second compact structure is found to be $E_{\text{ECEPP/3}} = -484$~kcal/mol 
while both compact structure have similar solvation energy, $E_{solv} = -178$~kcal/mol for 
the second compact structure. 
This may indicate shortcomings of our implicit solvent. This conjecture is supported by 
Ref.~\onlinecite{Garcia} where native-like configurations are observed as free energy
 global minimum  in simulations with an explicit solvent.

Note that the triangle-shaped compact configuration (``structure II") in Fig.~\ref{structures_b} 
has a total energy that is $\approx 30$~kcal/mol lower than the lowest-energy configuration found
in previous simulations Ref.~\onlinecite{KH05a} with the same force field.  
These simulations reported a lowest-energy  configuration where the middle helix is broken 
at residue {\it Gly21}, i.e. the configuration is built out of four helical segments. Similar
configurations are also observed in our simulations and are found to have energies comparable 
to the ones found previously in Ref.~\onlinecite{KH05a}.  Depending on the arrangement of the 
helices they either resemble the triangle shaped structure of type ``II'' or the other compact
structure of of type ``III'', and are grouped together with those structures  in our analysis of the distribution 
of varying structures as a function of temperature. 

\begin{figure}
\includegraphics[width=\columnwidth]{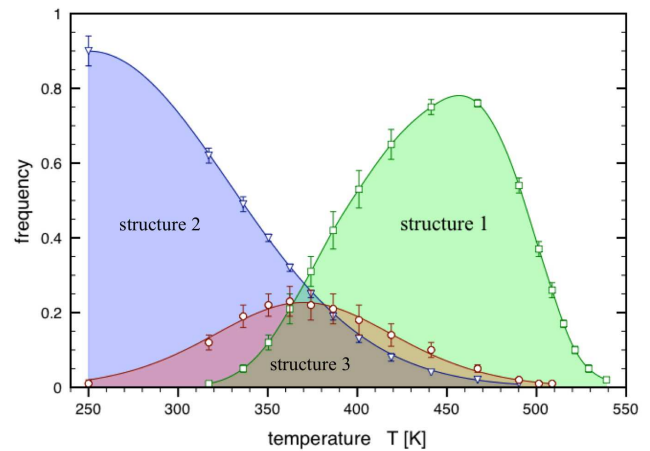}
\caption{\label{Distr} Frequency of the three dominating helical low-energy structures 
                        as function of temperature.}
 \end{figure}

Finally, we note that all three structures are built out of helical segments and therefore only 
observed with substantial frequency at and below the helix-coil transition at $T_c$. 
The extended structure (type I) is most frequent at temperatures between $441-467$~K.  
Its probability decreases with temperature and at $T=317$~K only about 1\% of the configurations 
belong to type I. Compact configurations (type II and III) apear with a frequency of more than 1\% 
only for temperatures below $T=490$~K. The native-like configurations of type III are slightly more frequent than the triangle-shaped of type II for a small temperature window above $T=362$~K where their frequency reaches with 23\% a maximum, see Fig.~\ref{Distr}. 
Further decreasing the temperature the frequency of configurations of type III diminishes while 
configurations of type II become more frequent. At $T=317$~K only 12\% of the configurations are
native-like, but already 62\% belong to type II. While the entropy of the triangle shaped structure 
(type II) is certainly lower than the entropy of the native structure (which might explain the suppression
of type-II structures at higher temperatures), the entropic gain of the native-like structure is compensated by an energetic gain of the maximally helical triangle structure at lower temperatures making the
triangle shaped structure the predominant one.
This observation clearly reflects a bias in the ECEPP/3 energy function toward helical structures. 
Growth of helices is energetically favored which in turn strongly  restricts the possible topologies the
folded structures can assume. This leads to the observed dominance of structures that while less 
compact and larger solvent accessible surface are are more helical than the one observed in 
experiment. 
Note that such a thermodynamic bias has not  been observed in Ref.~\onlinecite{Unres} where 
also configurations similar to the ones in Fig.~\ref{structures_a}-\ref{structures_c} were observed
during the simulation. This indicates that this coarse-grained model employed in Ref.~\onlinecite{Unres} (and the one of Ref.~\onlinecite{Irbaeck}) captures the effective interactions in protein A better than our all-atom energy function.
Hence, we should contemplate possible corrections to our energy functions which
should decrease the helix-forming tendencies. We are currently exploring this conjecture with a 
variant of the ECEPP force field proposed by Abagyan and co-workers \cite{Abagyan}.

\section{Conclusions}
We have performed parallel tempering simulations of the 46-residue B-fragment of protein A 
with an optimized temperature distribution and high statistics.
Our goal was to 
test the limitations set on protein simulations by our energy function, the ECEPP/3
force field with an implicit solvent. 
While we find native-like structures
with $\approx$ 3~\AA\  rmsd, these structures appear at room temperature with only 
$\approx 10$\% probability. 
Energetically favored are compact structures that are maximal helical, and more 
exposed to the surrounding solvent while less compact than the native structure.
This observation suggests the need for correction terms to the ECEPP/3 force field
which decrease the helix-forming bias when using the force field in combination 
with an implicit solvent.

{\em Acknowledgments --}
This work is supported, in part,  by a
research grant (CHE-0313618) of the National Science Foundation (USA).


\end{document}